\documentclass{aa}  

\usepackage{graphicx}
\usepackage{txfonts}
\usepackage{xcolor}
\begin{document}

   \title{Properties of Barred Galaxies with the Environment:}

   \subtitle{II. The case of the Cosmic Web around the Virgo cluster}

\author{Virginia Cuomo \inst{1}, J. Alfonso L. Aguerri 
          \inst{2,3}, Lorenzo Morelli \inst{4}, Nelvy Choque-Challapa \inst{4,5}, \and Stefano Zarattini \inst{6} %\fnmsep
          %\thanks{Just to show the usage
          %of the elements in the author field}
          }

   \institute{Departamento de Astronom\'ia, Universidad de La Serena, Av. Ra\'ul Bitr\'an 1305, 1700000, La Serena, Chile \\
   \email{virginia.cuomo@userena.cl}
   \and Instituto de Astrof\'{\i}sica de Canarias. C/ V\'{\i}a L\'actea s/n, 38200, La Laguna, Tenerife, Spain
   \and Departamento de Astrof\'{\i}sica de la Universidad de La Laguna, E-38206, La Laguna, Spain
   \and Instituto de Astronom\'{\i}a y Ciencias Planetarias, Universidad de Atacama, Avenida Copayapu 485, 1530000, Copiap\'o, Chile
   \and Departamento de Física, Universidad Técnica Federico Santa Maria, Av. Vicuña Mackenna 3939, 8940897, San Joaquín, Santiago, Chile
   \and Centro de Estudios de Física del Cosmos de Aragón (CEFCA), Unidad Asociada al CSIC, Plaza San Juan 1, 44001, Teruel, Spain
        }

   \date{\today}

% \abstract{}{}{}{}{} 
% 5 {} token are mandatory
 
  \abstract
   {Bars are elongated structures developed by a large fraction of disk galaxies in their central few kiloparsecs. However, the bar formation process is still not fully understood, particularly the role played by the galaxy environment in the formation and evolution of these structures.}
   {The aim of this work is to establish how the galaxy environment affects the evolution of bars by analyzing the bar structural parameters in a sample of galaxies located in three different galaxy environments: in the Virgo cluster, in filaments in the Cosmic Web around it, and in the field.}
   {We performed structural analysis using optical imaging from the DESI Legacy survey, measuring bar radii and disk scale lengths through Fourier analysis and surface brightness fitting techniques.}
   {After defining a homogeneous sample of barred galaxies across the three different galaxy environments in terms of color and magnitude, the median bar radii were found to be $2.54 \pm 0.34$ kpc, $3.29 \pm 0.38$ kpc, and $4.44 \pm 0.81$ kpc in the cluster, filaments, and field environments, respectively. In addition, the median bar radii scaled by the disk scale lengths was found to be $1.26\pm0.09$, $1.72\pm0.11$, and $2.57\pm0.21$ in the cluster, filaments, and field environments, respectively.  These results indicate that the galaxy environment has a significant influence on the structural parameters of bars, with bars in high-density environments being shorter and less prominent than those in the field.}
   {Our findings can be interpreted in terms of a slowing of the secular evolution of bars in dense galaxy environments. Barred galaxies located in clusters could experience a reduced rate of bar secular evolution due to various physical processes that occur in high-density environments, such as gas stripping, strangulation, or tidal interactions.}

   \keywords{galaxies: structure -- galaxies: formation -- galaxies: evolution -- galaxies: clusters: Virgo -- galaxies: Cosmic Web -- Cosmology: large-scale structure of the Universe}

\authorrunning{V. Cuomo et al.}
\titlerunning{Bar properties in different galaxy environments}
   \maketitle
%
%-------------------------------------------------------------------

\section{Introduction}

Stellar bars are common elongated structures observed in a significant fraction of galaxies across a wide range of masses and morphological types, including our own Milky Way \citep[see e.g.][]{Marinova2007, Aguerri2009, Wegg2015, Blandhawthorn2016}. These structures play a fundamental role in the secular evolution of galactic disks. They facilitate the exchange of angular momentum between the disk and the dark matter halo \citep[][]{Debattista2000, Athanassoula2003, Athanassoula2013}, channel gas towards the central regions—thereby enhancing star formation at the center or or quenching it within the bar regions of galaxies \citep[][]{Aguerri1999, Sanchezblazquez2011, Wang2012, Gavazzi2015} and potentially feeding supermassive black holes \citep[][]{Knapen2000, Corsini2003, Cheung2015} —and drive stellar migration across the disk, leading to the radial mixing of stellar populations \citep[][]{Minchev2010, DiMatteo2013}.

Numerical simulations of isolated disk galaxies suggest that the formation and evolution of stellar bars occur in three main phases. First, the bar forms rapidly within the initial $\sim$1–1.5 Gyr of evolution \citep[see e.g.][]{Athanassoula2002, MartinezValpuesta2006}. This initial growth is followed by a buckling phase, during which the bar thickens vertically and temporarily weakens in strength \citep[][]{Raha1991, MartinezValpuesta2004}. Finally, over several gigayears, the bar evolves secularly through the continuous exchange of angular momentum with other galactic components, particularly the dark matter halo \citep[][]{Debattista2000, Athanassoula2003}. This three-phase evolution is expected to hold for isolated disks, as highlighted by simulation studies specifically addressing this scenario. However, this formation process and the consequent evolution can be influenced by the galaxy environment in which the host galaxy lives \citep[][]{Peschken2019, Zana2019}.

Galaxies evolve within a wide range of galaxy environments over cosmic time. In low-density regions such as cosmic voids, galaxies tend to evolve in relative isolation, where the evolution is primarily driven by internal secular processes with minimal external influence. In contrast, in denser environments like large-scale filaments and galaxy clusters, galaxies are more likely to experience environmental effects such as tidal interactions, mergers, and ram-pressure stripping \citep[][]{Moore1996, Poggianti2017}. As galaxies migrate from voids through filaments toward the dense cores of clusters, the environment becomes increasingly influential, enhancing mechanisms that remove gas and suppress star formation, ultimately quenching their stellar populations \citep[][]{Peng2010, Kraljic2018, Zarattini2025}.

The diversity of galaxy environments can influence the formation and properties of stellar bars. A first explored aspect is related to the fraction of barred galaxies identified in different galaxy environments. Some studies reported an enhanced bar fraction in high-density environments compared to low-density ones \citep[][]{Thompson1981, Skibba2012}, which may hold specifically for lenticular galaxies \citep{barway2011,Lansbury2014}. However, other works have not found significant environmental dependence of the bar fraction \citep[][]{Eskridge2000, vanDenBergh2002, Li2009, Aguerri2009, Cameron2010, MendezAbreu2010, Marinova2012, Aguerri2023}.

One of the most comprehensive studies on the role of environment in bar formation was conducted by \cite{Lin2014}, who analyzed a sample of $\sim$ 30,000 galaxies from the Sloan Digital Sky Survey \citep[SDSS-DR7,][]{Abazajian2009}. They found that barred early-type galaxies are more strongly clustered than their unbarred counterparts on scales ranging from a few kiloparsecs to 1 Mpc, suggesting a connection between environment and bar formation, which holds exclusively for earlier morphological types. In turn, barred galaxies with a late-type morphology are associated with fewer neighbors within ∼50 kpc, implying that those galaxies may experience tidal forces from close companions which are able to suppress the formation/growth of bars.

Galaxies are distributed along the Cosmic Web, a hierarchical structure of interconnected filaments and walls, separated by giant voids, representing a variegate and complex type of galactic environment (e.g. \citealt{Peebles2001,Rojas2004,Kreckel2012,Dominguez-gomez2022}). Thanks to large photometric and spectroscopic surveys (e.g. SDSS, \citealt{Abdurro'uf2022}; CALIFA, \citealt{Sanchez2012}; MaNGA, \citealt{Bundy2015}; COSMOS, \citealt{Scoville2007}), it was possible to explore the Cosmic Web in details, discovering how dense environments such as galaxy clusters have an impact on the evolutionary path of galaxies and that their influence is effective up to several virial radii from the cluster center (e.g. \citealt{Tasca2009, Carollo2013,Cornwell2023}). 

In the nearby Universe, \citet{Castignani2022} investigated the influence of the large-scale structure on bar formation and evolution by analyzing the bar fraction across different environments, including galaxy cluster (in this case, Virgo), filaments, and the field. Their results revealed a slight decrease in the bar fraction from high-density environments, such as clusters, towards lower-density regions like the field, suggesting that the large scale environment plays a subtle but measurable role in the occurrence of bars.
 
Numerical simulations have shown that the properties of bars formed through galaxy interactions depend strongly on the strength of the perturbation. For instance, simulations of mild encounters, such as fly-by interactions, suggest that these events do not significantly alter the structural parameters of pre-existing bars. When bars are induced in otherwise stable disks by such interactions, they tend to develop over longer timescales and are generally weaker than those formed in isolation \citep[see][]{MartinezValpuesta2017}. In contrast, stronger interactions—such as the infall of a galaxy into a massive cluster—can produce bars with markedly different characteristics (see e.g. \citealt{Cuomo2022}). In this context, \citet{Lokas2016} investigated bar formation in Milky Way–like galaxies falling into a Virgo-like cluster and found that the resulting bars are typically longer, stronger, and rotate more slowly than those formed in isolation. Furthermore, bars formed in galaxies that have lost a significant fraction of their mass due to strong tidal interactions appear more prominent than those in isolated disks \citep[][]{Aguerri2009b, lokas2021}. 

Recent cosmological simulations have shown that the present population of barred galaxies had already developed their bars by $z=1$, or have remained unbarred, depending on the galaxy environment in which they evolve \citep[][]{RosasGuevara2024}. In particular, galaxy interactions and mergers can lead to a variety of outcomes, including the formation or dissolution of bars, or even the complete disruption of the disk structure, which ultimately drives morphological transformations \citep[see e.g.,][]{RosasGuevara2024, Zheng2025}.

The aim of this paper is to investigate the influence of different galaxy environments on the structural properties of disk galaxies. It is well established that high-density environments, such as the Coma cluster, can significantly affect the structure of galaxy disks \citep[see][]{gutierrez2004, aguerri2004,MendezAbreu2012}. Since bars in galaxies trace the dynamical state of disks, they are valuable tools for probing environmental effects. However, the impact of the environment on bar properties remains a matter of debate. 

This paper is the second in a series investigating the influence of the galaxy environment on the structural properties of barred galaxies. In the first study \citep[][hereafter Paper I]{Aguerri2023}, we analyzed a sample of morphologically classified early-type barred galaxies located in the Virgo cluster. We measured their bar radii and compared them with those of barred galaxies of similar luminosity that reside in lower-density environments. Our results showed that bars in cluster galaxies tend to be physically shorter than those in galaxies of comparable luminosity located in less dense regions. However, this difference vanished when bar lengths were normalized by galaxy size, as cluster galaxies at fixed luminosity also exhibited smaller overall sizes than their field counterparts.
In the present work, we extend this analysis by including a larger and more diverse sample of galaxies spanning three distinct environments: the Virgo cluster, filaments in the Cosmic Web around Virgo, and the field. Unlike the previous study, here we match galaxies not only in luminosity, but also in stellar color and disk scale length, enabling a more robust assessment of how environment influences bar properties.

This paper is organized as follow: in Sect.~\ref{sec:sample} we describe the selection of the three galaxy samples corresponding to the different galaxy environments. In Sect.~\ref{sec:results} we present our measurements to characterize the properties of bars and host disks, while in Sect.~\ref{sec:discussion} we discuss the implications of our results. Our conclusions are reported in Sect.~\ref{sec:conclusions}. As done in Paper I, the cosmology adopted for this work is $H_0 = 70$ km s$^{-1}$ Mpc$^{-1}$, $\Omega_m = 0.3$, and $\Omega_\Lambda = 0.7$.

%--------------------------------------------------------------------
\section{Sample of barred galaxies}
\label{sec:sample}

This section describes the selection process of barred galaxies across the different environments considered in this study, detailing the data sources, selection criteria, and methods used to assign galaxies to their respective environments. We selected barred galaxies hosted in the Virgo cluster, in the filaments building the Cosmic Web around Virgo, and in the field, using different catalogs and techniques. First, we used the \citet{Castignani2022}'s catalog, which provides cluster and filament membership; then, the morphological classification provided by the Galaxy Zoo DESI project \citep{Walmsley2023}, which allows to identify barred and unbarred galaxies; and finally, we computed the local environment following the methodology of \cite{Zarattini2025} to distinguish between the three different environments. The final result of this selection is a sample of morphologically identified barred galaxies belonging to the three distinct environments.

\begin{figure}
    \centering
    \includegraphics[scale=0.10]{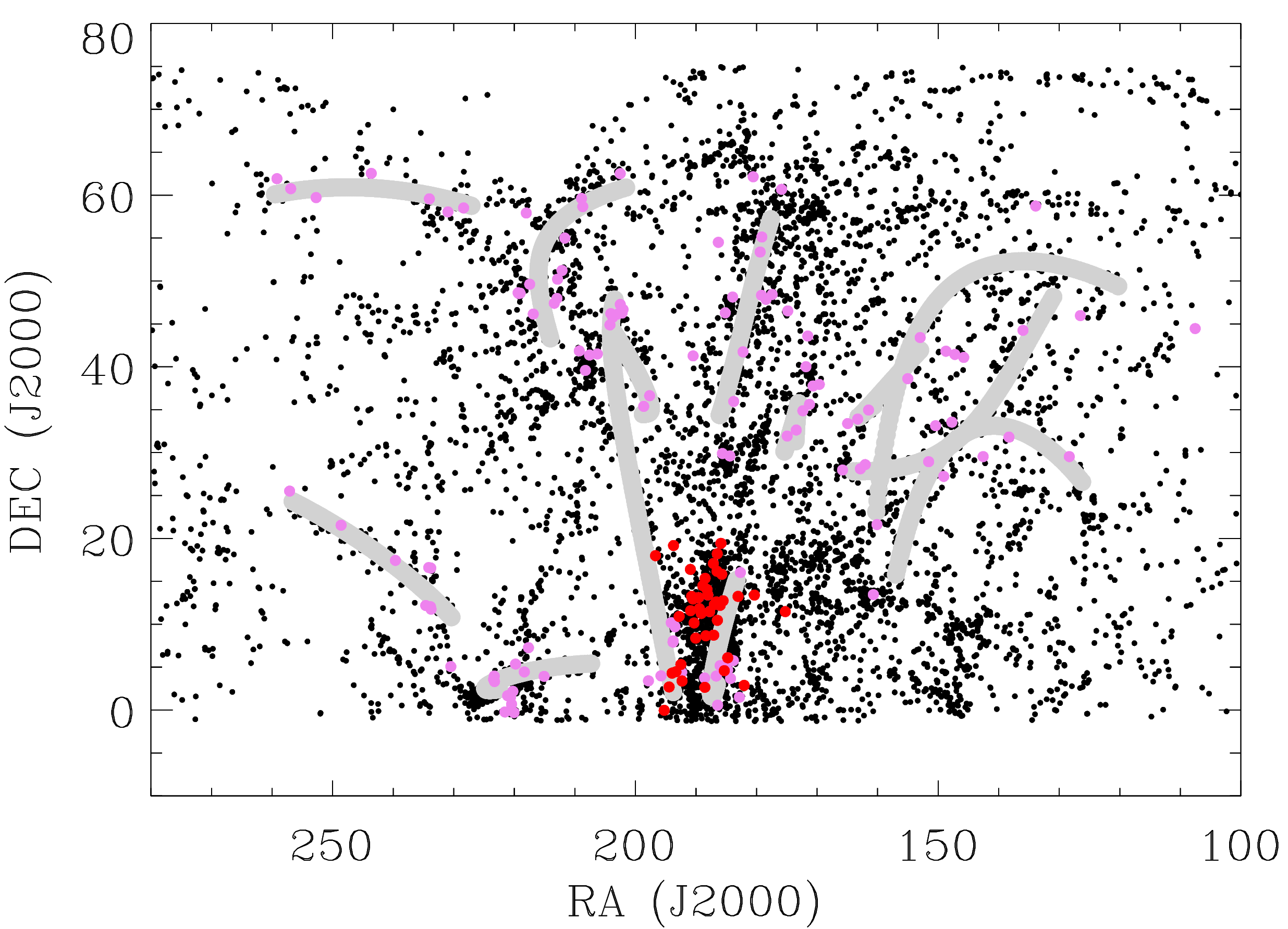}
    \caption{Galaxies in Virgo infall region (black points). In particular, the grey lines mark the filaments identified by \cite{Castignani2022}. Our selection of barred galaxies located in the cluster and filaments are highlighted (red and violet points, respectively).}
    \label{fig:virgo_infall}
\end{figure}

\subsection{Barred galaxies from different galaxy environments}

We first identified galaxies hosted in the Virgo cluster and in the filaments building the Cosmic Web around the Virgo cluster, using the \citet{Castignani2022}'s catalog. Focusing on the Virgo infall region, this study includes both the Virgo cluster itself and a network of 13 surrounding filaments that define its infall structure. \citet{Castignani2022} provide environmental classifications for each galaxy, indicating whether it belongs to the cluster or to one of the filaments. As a first step, we selected all the galaxies identified as residing within these two environments.

To determine the morphological types and identify barred galaxies within this selection, we used the morphological classification provided by the Galaxy Zoo DESI project \citep{Walmsley2023}, aimed at offering detailed morphological information, including the presence of bars, of more than 8 million galaxies in the DESI Legacy Imaging Surveys \citep{Dey2019}, based on automated measurements made by deep learning models trained on citizen science votes. We considered as barred those galaxies from the \citet{Castignani2022}'s catalog that have a marked \textit{Strong bar probability} in the Galaxy Zoo DESI morphological catalog \citep[see][]{Walmsley2023}. This classification investigates whether each galaxy hosts a strong, weak or no bar, providing a vote fraction for the corresponding \textit{Bar Strength} parameters ($b_s$ for \textit{strong} bars, $b_w$ for \textit{weak} bars, and $b_{no}$ for \textit{no} bars). Therefore, the parameter $b_s$ quantifies the probability that a given galaxy would be classified as having a strong bar. Specifically, it represents the expected fraction of volunteer votes for the answer 'bar - strong', as predicted by the Zoobot model, a convolutional neural network trained on Galaxy Zoo classifications. The parameter takes values between 0 (no strong bar expected) and 1 (all votes expected to be strong bar). After several values used for this parameter and to maximize the identification of strongly barred galaxies in this preliminary selection, we decided to include galaxies with a vote fraction for the \textit{Bar Strength} parameters $b_s$ greater than 0.1. This first selection results in 74 and 131 galaxies in the cluster and filament environments, respectively. Figure \ref{fig:virgo_infall} shows the galaxies belonging to the Virgo infall region, adapted from \citet{Castignani2022}. The location of our selected barred galaxies in the cluster and filaments has been highlighted by colored points (red for barred galaxies in the cluster and violet for filaments).

Moreover, we defined a sample of barred galaxies from the field. Adopting the same classification included in the Galaxy Zoo DESI morphological catalog, we applied the cut in bar strength as done in our previous selections, and considered galaxies with redshift $z < 0.03$ to ensure sufficient spatial resolution for reliably characterizing bars. We take care that these barred galaxies were not included in \cite{Castignani2022}'s catalog, in order to confirm that the selected field galaxies do not belong to the Virgo cluster or its associated filaments.

\begin{figure}
    \centering
    \includegraphics[scale=0.5]{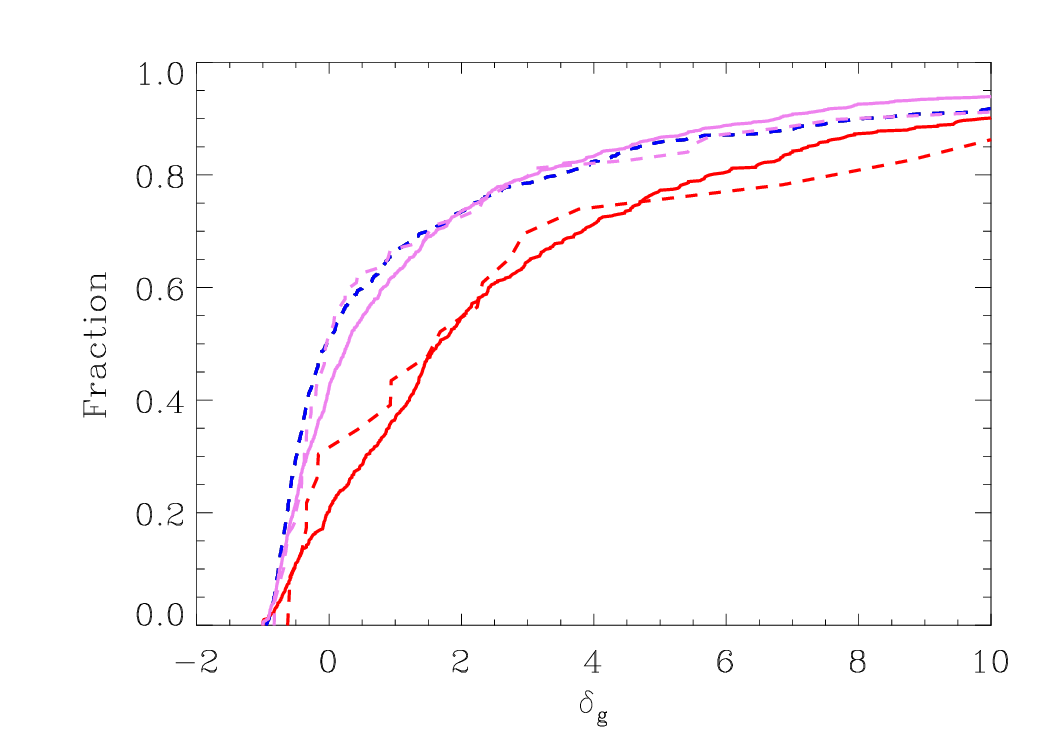}
    \caption{Cumulative distribution function of the local galaxy overdensity of different galaxy sub-samples defined in this work, calculated following \citet{Zarattini2025}. The solid lines show the behavior of all the cluster (red), and filament (violet) galaxies included in \cite{Castignani2022}'s work (barred or not). The dashed lines show the behavior of the morphologically selected barred galaxies, which have been obtained from \citet{Walmsley2023} and adopted in this study. The dashed red, violet, and blue lines mark the cluster, filament and field barred galaxies, respectively.}
    \label{fig:overdens_SDSS}
\end{figure}

We further characterized the different environments of the three galaxy samples (belonging to the cluster, filaments, and field) by exploring the local galaxy overdensity ($\delta_{g}$) for the galaxies in each environment. The local environment parametrized by $\delta_{g}$ was obtained from \cite{Zarattini2025}, who measured $\delta_{g}$ of all galaxies with a redshift estimate included in the SDSS-DR16 catalog \citep[][]{Ahumada2020}, building their own catalog of galaxy overdensities. They obtained the projected local galaxy density ($\Sigma_{g}$) around each galaxy in the SDSS sample by identifying the five nearest neighbors within a velocity range of $\pm 3000$ km s$^{-1}$. To account for spectroscopic incompleteness, the authors applied a photometric correction to $\Sigma_{g}$. The final local galaxy overdensity was defined as $\delta_{g} = (\Sigma_{g} - \Sigma_{z,g})/\Sigma_{z,g}$, where $\Sigma_{z,g}$ is the median galaxy density within a redshift slice of $\Delta z = 0.005$. We did a cross-match between our galaxy samples from the three different environments and the \citet{Zarattini2025}'s catalog. Figure~\ref{fig:overdens_SDSS} shows the local cumulative overdensity distributions for cluster and filament galaxies from the \cite{Castignani2022}'s catalog (red and violet solid lines) and our selected samples of barred galaxies by using the DESI morphology \citep[][]{Walmsley2023} from cluster, filaments and field (dashed-red, dashed-violet and dashed-blue lines). As expected, galaxies in the cluster sample are located in regions of higher local overdensity compared to those in the filament and field samples. The barred galaxies in the cluster and filament environments exhibit overdensity distributions that closely follow those of their respective parent samples.

Interestingly, the barred field galaxies show a local overdensity distribution similar to that of galaxies in filaments. This suggests that distinguishing between pure field and filament environments based solely on local overdensity is challenging. To address this, we selected a subsample of barred galaxies with the lowest overdensities ($\delta_{g} < -0.25$) to define the field sample, and checked that each selected galaxy was not included in \cite{Castignani2022}'s catalog. This subsample, consisting of 177 barred galaxies, represents our final selection of a galaxy sample from the field and has been used to compare bar properties in the most under-dense environments with those in higher-density environments. We refer to this galaxy sample as field one in the following.

{It is important to note that the galaxy overdensity computed by \citet{Zarattini2025} is based on projected measurements, which could introduce potential biases when adopted to classify galaxy environments. Indeed, projection effects may lead to the inclusion of galaxies that are not physically associated, particularly along the line of sight, to a given environment, and consequently dilute the intrinsic differences between field, filament, and cluster regions. Moreover, the choice of the redshift window used to select neighbors further affects the balance between contamination by interlopers and the exclusion of genuine members with significant peculiar velocities. These effects could be especially relevant for filamentary structures, where elongation along the line of sight enhances projection uncertainties. These projection effects have been analyzed in the literature, including the smearing of the galaxy distribution due to redshift uncertainties \citep[][]{Cooper2005}, as well as their impact on the inferred properties of galaxy clusters \citep[see e.g.,][]{Wojtak2018, Sunayama2020}. As a result, the environmental trends reported here should be interpreted with these limitations in mind.}

\subsection{Quantitative identification of bars}
\label{sec:fourier}

Following the approach presented in Paper I, we adopted a Fourier analysis to quantitatively confirm whether the selected barred galaxies from different environments actually host bars. 

We used publicly available co-added $r$-band images from the DESI Legacy survey and followed the approach described in Paper I. In summary, we first performed an isophotal fitting analysis using the ELLIPSE task \citep[][]{Jedrzejewski1987} from the IRAF\footnote{IRAF is distributed by the National Optical Astronomy Observatory, which is operated by the Association of Universities for Research in Astronomy (AURA) under a cooperative agreement with the National Science Foundation} package to derive the disk ellipticity (and consequent disk inclination $i$) and position angle (PA) from the outer isophotes (corresponding to the disk-dominated region). The derived ellipticity and PA were used to deproject the $r$-band images, by stretching the images of the galaxy along the disk's minor axis by a factor equal to $1/\cos i$, and conserving the flux. Next, we decomposed the azimuthal surface brightness profile $I(r,\phi)$\footnote{Assuming polar coordinates in the galaxy disk $(r,\phi)$.} of each sample galaxy extracted by the deprojected images, into a Fourier series: 

\begin{equation}
I(r, \phi) = \frac{A_0 (r)}{2} + \sum A_m (r) \cos(m\phi) + \sum B_m (r) \sin(m\phi)
\end{equation}

The radial profiles of the amplitudes of the Fourier components $I_m(r) = (A_m^2(r) + B_m^2(r))^{1/2}$ are derived to define the relationship between the amplitude of the even Fourier mode $m = 2,4,6$ and the zeroth mode $m = 0$, corresponding to the total light measured at each radial annulus, allowing us to identify and characterize the bars. In barred galaxies, the even Fourier modes (in particular the $m=2$ component) are significantly stronger than the odd modes within the bar region (for more details, see Paper I and references therein). 

To quantitatively identify a bar, we adopted a value of $(I_2/I_0)_{\rm max}=0.2$ as minimum threshold (which corresponds to a minimum value for the bar strength $A_2=(I_2/I_0)_{\rm max}$, as consistently done in several studies based on similar analyzes, see e.g. \citealt{Aguerri2003, Cuomo2019, RosasGuevara2024}). Those galaxies with bar strength $A_2 < 0.2$ were excluded from the samples.

The Fourier analysis allowed us to confirm the presence of bars in the visually defined samples of barred galaxies from different environments presented in the previous sub-section, and to exclude some galaxies erroneously identified as barred by \cite{Walmsley2023}, after revising both the results of our Fourier analysis and conducting a visual inspection of the images. Our final samples of quantitatively confirmed barred galaxies include 43 objects in the Virgo cluster, 116 in filaments, and 93 from the field. After applying our Fourier-based quantitative classification followed by visual inspection, the preliminary selection from \cite{Walmsley2023} was reduced by about 40\% for the Virgo sample, 11\% for the filaments and nearly 50\% for the field. This two-step approach, combining quantitative Fourier analysis with visual inspection, ensures a robust and conservative identification of strongly barred galaxies across different environments. In turn, the adopted restrict threshold to identify barred galaxies from \cite{Walmsley2023}'s catalog ensures a reliable selection of strongly barred galaxies, which gives us confidence that most of these objects, if not all, have been included in our preliminary selection.

%\textcolor{red}{For Virginia: We have to add a phrase indicating the minimum max(A2) adopted to select a galaxy as barred one.}

%\textcolor{red}{For Alfonso and Lorenzo: do we want to add a Figure about the Fourier analysis? This was already included in Paper I, so I would say no.}

\subsection{Building a homogeneous sample of barred galaxies}

It is well established that both the environment and the mass of the galaxy influence the morphological type, the rate of star formation, and the evolution of galaxies \citep[e.g.,][]{Dressler1980}. Therefore, in order to identify possible environmental and mass effects on the evolution of bar properties, we considered necessary to define a sample of barred galaxies with similar characteristics in terms of mass dependencies: namely, color and magnitude.

\begin{figure}
    \centering
    \includegraphics[scale=0.155]{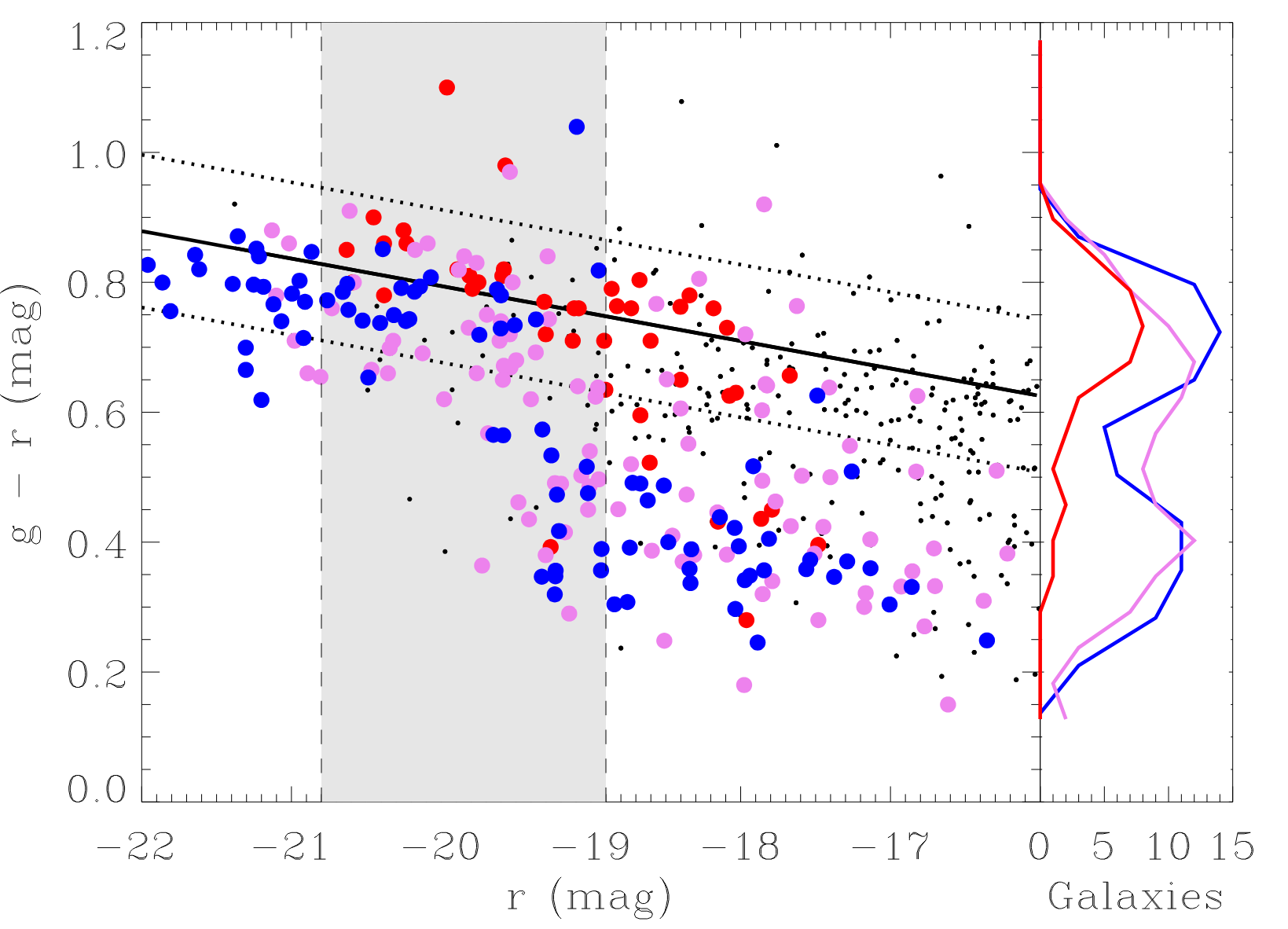}
    \caption{Left panel: Color–magnitude diagram of barred galaxies located in the Virgo cluster (red points), filaments (violet points), and field (blue points), selected as described in Sec.~\ref{sec:fourier}. Non-barred galaxies hosted in the Virgo cluster from \citet{Castignani2022} are marked as well (black points). The solid line shows the best fit to the red sequence of Virgo cluster galaxies, while the dotted lines indicate the $1\sigma$ uncertainty of the fit.  {The shaded grey area represents the selected luminosity range of the galaxies.} Right panel: $g-r$ stellar color distribution of the galaxies located in the cluster (red line), filament (violet line) and field (blue line).}
    \label{fig:color_mag}
\end{figure}

Figure \ref{fig:color_mag} displays the color–magnitude diagram of the selected barred galaxies residing in the three different galaxy environments considered in this study. The selected barred galaxies span a wide range of $r$-band absolute magnitudes, from bright to dwarf systems, and are distributed across the diagram from the red sequence to the blue cloud, indicating a diversity of morphological types. We fitted the red sequence (marked by a solid black line) using all galaxies (barred and non-barred) in the Virgo cluster from the \citet{Castignani2022}'s catalog. Barred galaxies in the cluster environment (red points in Fig. \ref{fig:color_mag}) are primarily concentrated within $\pm1\sigma$ of the red sequence fit (marked by dotted-black lines) and are massive ($M_r < -19.5$), with only a few fainter systems located in the blue cloud. In contrast, barred galaxies in filament and field environments exhibit a bimodal color distribution: they lie along the red sequence for $M_r < -19.5$ and predominantly populate the blue cloud at fainter magnitudes ($M_r > -19.5$), {a trend consistent with the morphology–density relation, where galaxies in denser environments preferentially lie on the red sequence while fainter galaxies more often populate the blue cloud (Dressler 1980).}

Bar properties are known to correlate with the morphological type of galaxies. In particular, longer and more prominent bars are typically found in early-type galaxies \citep[see][]{Erwin2005, Aguerri2009, Gadotti2011, Cuomo2020}. To minimize the impact of this morphological dependence, we restricted our analysis to barred galaxies redder than $1\sigma$ of the fitted red sequence in the color–magnitude diagram for the three galaxy environments, which are expected to be dominated by early-type morphologies\footnote{{Throughout this paper, when referring to “early-type barred galaxies” we mean disk galaxies of early Hubble types (S0 and Sa–Sb).}}.

In addition, bar properties also strongly correlate with the stellar mass or luminosity of galaxies: more massive or luminous galaxies tend to host longer bars \citep{Cuomo2020}. As shown in Fig.~\ref{fig:color_mag}, the luminosity distributions of galaxies within the red sequence differ among the cluster, filament, and field samples. To control for this effect, we selected galaxies within a fixed luminosity range (marked by vertical dashed-black lines), $-20.8 < M_r < -18.9$, which is well populated across all three environments. The final selection is marked by the (vertical and inclined) dashed and dotted lines in Fig.~\ref{fig:color_mag}.

The final three selected samples based on galaxy color and luminosity consist on 70 galaxies located in Virgo cluster (22), filaments (26) and field (22) environments (crf. Fig.~\ref{fig:color_mag}). These galaxies are representative of Milky Way–like systems in terms of stellar mass and are expected to correspond to early-type disk morphologies. Table \ref{tab:gal_prop} shows the median $g-r$ stellar colous and absolute $r-$band magnitudes of the galaxies of the different samples

\begin{table}[]
    \centering
    \caption{Median values and dispersions of the $g-r$ stellar color and $r-$band absolute magnitudes of the galaxies in the different environments.}
    \begin{tabular}{lccc}
    \hline \hline
    Parameter & Virgo cluster & filaments & field \\
    \hline
    $\langle g-r\rangle$ & $0.81\pm0.02$ & $0.74\pm0.02$ & $0.78\pm0.01$ \\
    %$\langle S_{\rm b}\rangle$ & $0.4\pm0.2$ & $0.5\pm0.2$ \\
    $\langle M_{r} \rangle$ & $-19.75\pm0.12$ & $-19.61\pm0.09$ & $-20.18\pm0.11$ \\
    \hline
    \end{tabular}
   \tablefoot{Uncertainties of the median values correspond to $\sigma/\sqrt{N}$, where $\sigma$ and $N$ are the standard deviation and the number of galaxies.} 
    \label{tab:gal_prop}
\end{table}

%\subsection{Barred galaxies from the Cosmic Web}

%\subsection{Barred galaxies from field}

%Explain why we redefined the field. Figure about overdensity?

%\subsection{Galaxy Densities for the sub-samples}

\subsection{Additional sanity tests}

To doule-check the presence of bar-like structures in our selected galaxies, we performed some additional tests. The first test involved obtaining the visual morphological classification of the galaxies. There is no homogeneous visual morphological classification available from a single source for all galaxies in the sample. Therefore, we collected classifications from two different databases: Hyper-Linked Extragalactic Database \citep[HYPERLEDA, ][]{Makarov2014} and NASA/IPAC Extragalactic Database (NED; \url{https://ned.ipac.caltech.edu}). Most of the cluster (86\%) and filament (72\%) samples were classified as SB or SAB types, while this fraction was smaller (18\%) for the field sample.

In addition to the morphological classification, we confirmed the presence of bars by inspecting the images directly. Iso-light contours were produced on the DESI images to identify elongated bar-like structures in the central regions of the galaxies. This test led to the exclusion of 9 galaxies from the sample, due to strong dust extinction in their central regions or because they were highly inclined, preventing clear bar detection and biasing the Fourier analysis. For the remainning galaxies clear bar-like structures appear in the central regions of the galaxies.

After these two consistency checks, our final sample of barred galaxies consists of 54 objects, distributed across cluster (20), filament (22), and field (12) environments.

%\begin{figure*}
%    \centering
%    \includegraphics[scale=0.45]{test_color_cut/color.png}
%    \includegraphics[scale=0.45]{test_color_cut/rbar_mag.png}
%    \includegraphics[scale=0.45]{test_color_cut/h_mag.png}
%    \includegraphics[scale=0.45]%{test_color_cut/scaled_rbar_mag.png}
%    \caption{Caption}
%    \label{fig:enter-label}
%\end{figure*}

\section{Results}
\label{sec:results}

In this section, we present the results related to the structural parameters of the barred galaxies analyzed in the three different galaxy environments. Specifically, we show the disk scale lengths, bar radii, and the ratio between the bar radius and the disk scale length of the galaxies.

\subsection{Disk structural parameter: the disk scale length}

\begin{figure*}
    \centering
    \includegraphics[scale=1.0]{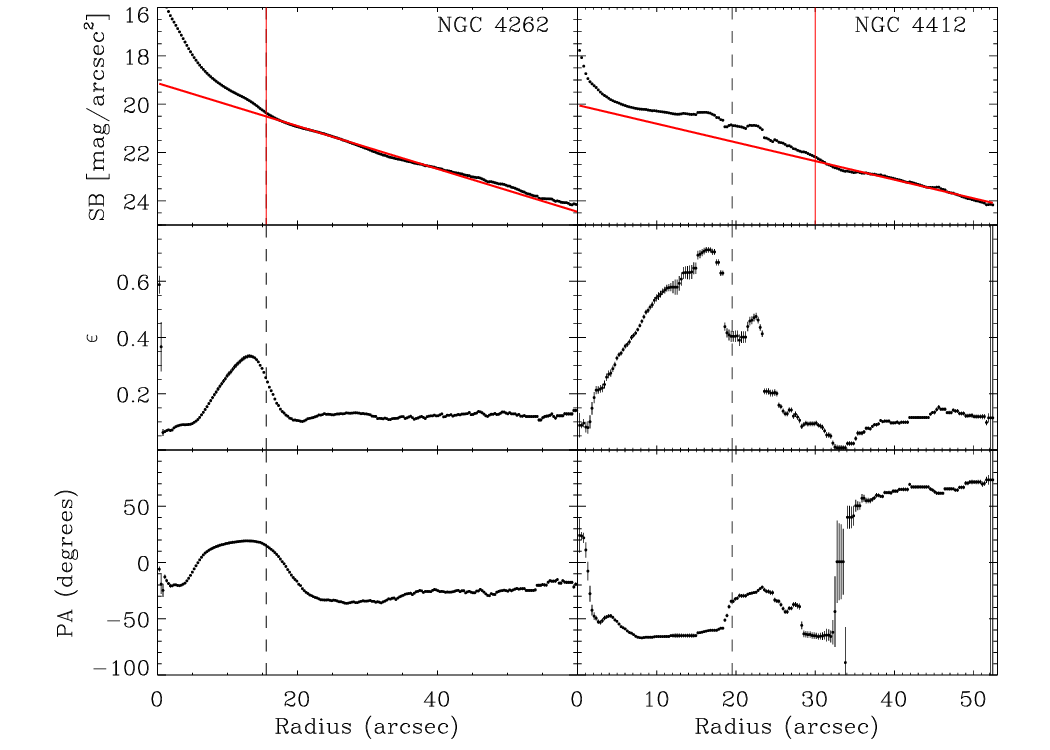}
    \caption{Two examples of {barred} galaxies showing their radial profiles of surface brightness (top panels), ellipticity (middle panels), and position angle (bottom panels). The vertical black dashed line indicates the bar radius of each galaxy, while the vertical solid red line marks the innermost radius used for the disk scale length fit (red line).}
    \label{fig:h_fit}
\end{figure*}

The surface brightness profiles of each galaxy in the three samples were obtained by fitting ellipses to their isophotes using the ELLIPSE task from the IRAF package, as described in Sect.~\ref{sec:fourier} (see Fig.~\ref{fig:h_fit} for some examples of our analysis).

Photometric decompositions of galaxy surface brightness profiles have shown that the outer regions of disk galaxies, where the disk component dominates, are well described by an exponential law \citep[see e.g.][]{Freeman1970, deJong1996, Prieto2001, Graham2001}. This mathematical expression is given by

\begin{equation}
    I(r) = I_{0} e^{-r/h}
\end{equation}

where $I_{0}$ and $h$ are the disk central surface brightness and scale length, respectively. 

We derived the disk scale lengths of our galaxy sample by fitting an exponential function to the outer regions of the surface brightness profiles of the galaxies. Specifically, we performed the fit at radii larger than the bar radius, measured by the Fourier analysis presented in Sect.~\ref{sec:fourier}. It is important to note that we did not carry out a full surface brightness decomposition across all radii, where the contributions of multiple galactic components would need to be considered simultaneously. Instead, we applied an exponential fit solely to the external regions, where the surface brightness is predominantly determined by the disk component. This approach allows for a good approximation of the disk scale, as demonstrated by \citet{Erwin2005}.

In most cases, we fitted the exponential law using all points of the surface brightness profile beyond the bar radius. However, in a few instances, the bar was sufficiently prominent to influence the surface brightness profile well beyond its measured radius. In those cases, the exponential fit to the disk was performed fixing the starting radius {beyond the end of the bar}, ensuring that the fitted disk profile was not significantly affected by the bar structure.

Adopting this approach, the extrapolated exponential profile of the disk lies below the total surface brightness profile when projected toward the galaxy center. Figure \ref{fig:h_fit} illustrates two examples of these profiles, corresponding to the typical cases {of barred galaxies} described above.

Figure \ref{fig:h_disc} shows the cumulative distributions of the disk scale lengths for galaxies in the cluster, filament, and field environments. There is a slight difference between the cumulative distribution of disk scale lengths for field galaxies compared to those in clusters and filaments. Specifically, the median disk scale length of field galaxies is marginally larger ($h_{\mathrm{field}} = 2.34\pm0.33$ kpc) than that of cluster and filament galaxies ($h_{\mathrm{cluster}}= 2.10\pm0.27$ kpc and $h_{\mathrm{filament}}= 2.17\pm0.35$ kpc, results are reported in Table~\ref{tab:mean_value}). We performed several statistical tests to assess whether the differences among the three distributions are statistically significant (results are summarized in Table~\ref{tab:test_stat}). Two of these tests focus on comparing the overall distribution functions: the Kolmogorov–Smirnov (KS) test and the Anderson–Darling (AD) test. The main difference between them is that the AD test gives more weight to the tails of the distributions, making it more sensitive to differences in the extremes. Both tests indicate that there are no statistically significant differences between the three distributions (we found {{$p_{KS}$ and $p_{AD}$ values} larger than 0.05), when comparing two samples at a time. Moreover, we applied the AD test for the three samples at once. This indicates the probability that the three different samples are part of the same distribution, and we found no differences. Additionally, we applied the Welch t-test to examine whether the distributions differ significantly in their median values. This test also yielded no significant differences. We therefore conclude that the disk scale length distributions for galaxies in the three environments considered do not show statistically significant differences, neither in their shape nor in their median values.

%However, this difference is not statistically significant. We performed a Kolmogorov-Smirnov (KS) test on the three cumulative distributions to verify if there are statistical differences between the distributions of the disk scale lengths. When comparing the three galaxy samples two at a time, the three KS probabilities were $p > 0.05$, indicating that the three distributions of disk scale lengths are consistent with being drawn from the same parent population.

The fact that the three disk scale length distributions are statistically similar indicates that we have three samples of disk galaxies located in different environments but with comparable disk properties. This suggests that potential differences observed in the bar parameters may be linked to environmental effects, given that the galaxy samples display broadly similar color/morphology, luminosity, and disc scale lengths (which, however, is the only structural parameter explicitly examined in this work).

\begin{figure}
    \centering
    \includegraphics[scale=0.53]{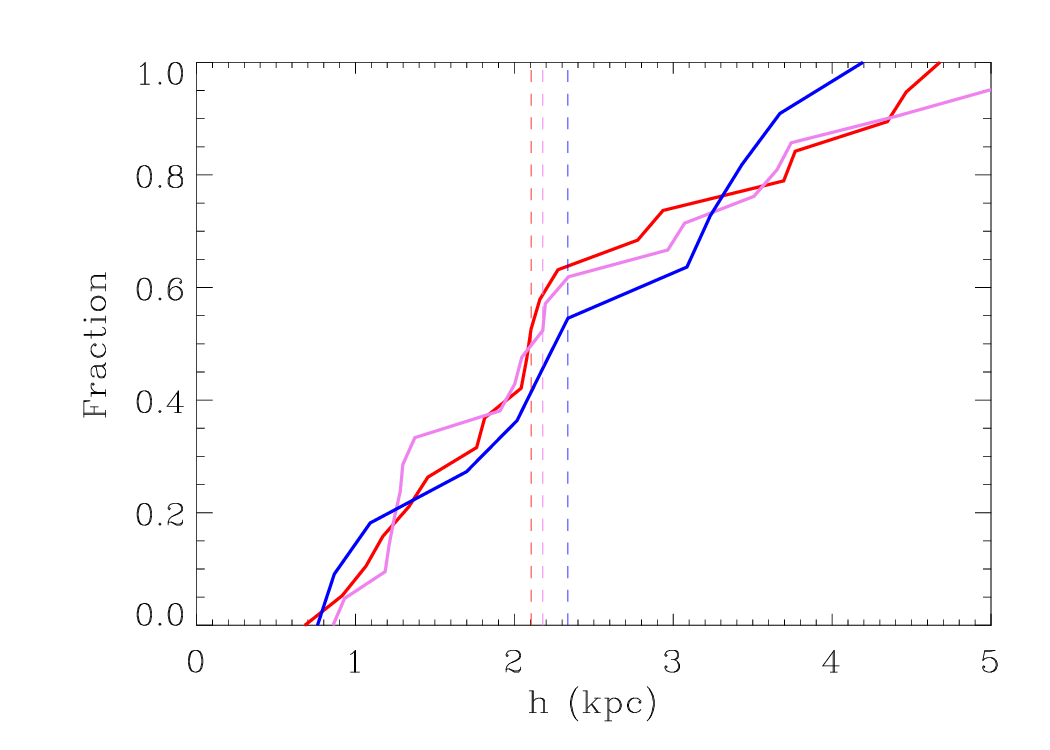}
    \caption{Disk scale lengths in physical units for galaxies located in the cluster (red line), filament (violet line), and field (blue line) environments. Vertical dashed lines shows the median values of the disk scale lengths for galaxies in the cluster (red), filament (violet) and field (blue). The vertical lines mark the median values of the disk scale length per each galaxy sample.}
    \label{fig:h_disc}
\end{figure}

\subsection{Bar structural parameter: the bar radius}

Following our previous works \citep[][]{Aguerri2000, Aguerri2015, Cuomo2019, Aguerri2023} we determined the bar radius ($R_{bar}$) by using the Fourier decomposition of the surface brightness of the galaxies of the samples, already described in Sect.~\ref{sec:fourier}. In particular, we adopted the method developed by \citet{Aguerri2000} to determine $R_{bar}$. This method uses the shape of the radial profile of the so-called bar-interbar intensity profile which is defined as a lineal combination of the even Fourier modes of the galaxy. Assuming that the bar-interbar intensity profile has a typical Gaussian shape for a barred galaxy, the bar radius is identified as the second furthest radius where the bar-interbar intensity profile reaches its median value between the minimum one, corresponding to the the center of the galaxy, and its maximum one, measured within the bar region (see Paper I for more details). 

\begin{figure}
    \centering
    \includegraphics[scale=0.53]{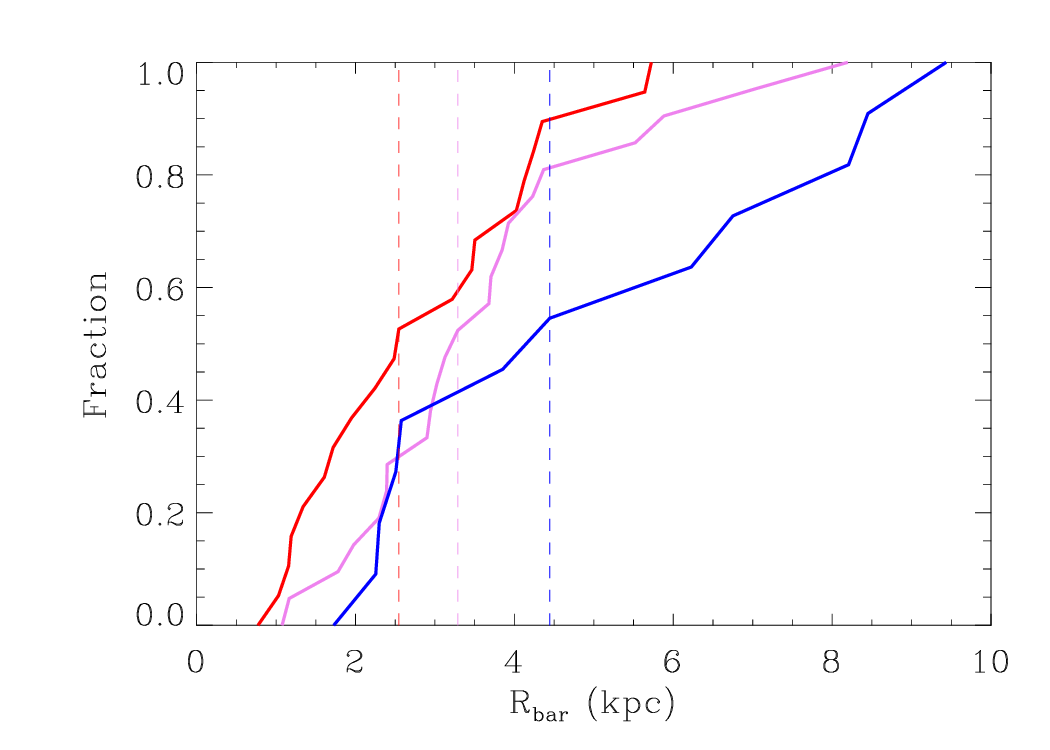}
    \caption{Bar radius in physical units for galaxies located in the cluster (red line), filament (violet line), and field (blue line) environments. Vertical dashed lines shows the median values of the bar radius for galaxies in the cluster (red), filament (violet) and field (blue). The vertical lines mark the median values of the bar radius per each galaxy sample.}
    \label{fig:rbar_kpc}
\end{figure}

Figure \ref{fig:rbar_kpc} shows the cumulative distribution functions of the derived bar radii for the cluster, filament, and field galaxy samples. Cluster and field galaxies exhibit differences in their bar sizes: in particular, cluster galaxies tend to host shorter bars compared to similar galaxies in the field. The median bar radii for cluster, filaments and field galaxies are reported in Table~\ref{tab:mean_value}. There is a trend in bar radius indicating that bars in cluster galaxies are shorter than those in the field, with the median values for filament galaxies lying in between. The KS test suggests that the three $R_{\mathrm{bar}}$ distributions are statistically similar. In contrast, the AD test reveals statistically significant differences between the three environments. In particular, differences are observed when the three samples are compared at once, and between bar radii from Virgo and the field. Similarly to the latter test, the Welch t-test show that the distributions $R_{\mathrm{bar}}$ in the Virgo and field samples differ significantly in their median values, while no significant differences are found in the other pairwise comparisons. We therefore conclude that bars in cluster and field galaxies are statistically different both in the shape and in the median values of their $R_{\mathrm{bar}}$ distributions (we consider that the results from the KS test can be partially biased by the small numbers of the three samples). See Tab. \ref{tab:test_stat} for the values of the probabilities of the statistical tests.

%However, the KS test shows that the bar radius distributions of cluster and field galaxies are not statistically different ($p_{KS}=0.14$). Larger number of galaxies in the cluster and specially in the field samples would be needed to confirm statistically the difference between the bar radius between these two samples. The median bar radius of filament galaxies is $R_{\mathrm{bar,filament}}=3.29\pm0.38$ kpc, which lies between the cluster and field median values; according to the KS test, the bar radius distributions of filament galaxies are not statistically different from either the cluster ($p=0.38$) or field ($p=0.32$) distributions.

%A similar result was reported by \citet{Aguerri2023}, who found that bars in the Virgo cluster were significantly shorter than those in their corresponding field sample. Specifically, they measured a mean bar radius of $R_{\mathrm{bar}} = 2.6 \pm 1.5$ kpc for the Virgo cluster galaxies, a value very similar to that obtained in the present work. However, the physical bar size in their field sample was $R_{\mathrm{bar}}=6.1 \pm 3.0$ kpc, notably larger than the value measured for our field galaxies (results are reported in Table~\ref{tab:mean_value}). These differences in the bar lengths of the field samples likely reflect differences in the selection criteria for field galaxies between \citet{Aguerri2023} and this study, suggesting that selection effects in the field sample of \citet{Aguerri2023} could explain the discrepancy.

\subsection{Bar structural parameter: bar radius scaled by the disk scale length}

We scaled the bar radius of the galaxies by their disk scale lengths, computing the ratio $R_{\mathrm{bar}}/h$. 
{Although the disk scale-length distributions in our different environments are statistically similar, we  present $R_{\mathrm{bar}}/h$ because this is a standard approach in the literature and allows us to account for galaxy-to-galaxy variations in disk size, providing a more robust comparison of relative bar prominence.}

%This parameter quantifies the prominence of the bar relative to the disk: larger values of $R_{\mathrm{bar}}/h$ indicate that the bar occupies a greater fraction of the disk, resulting in more prominent bars.

\begin{figure}
    \centering
    \includegraphics[scale=0.53]{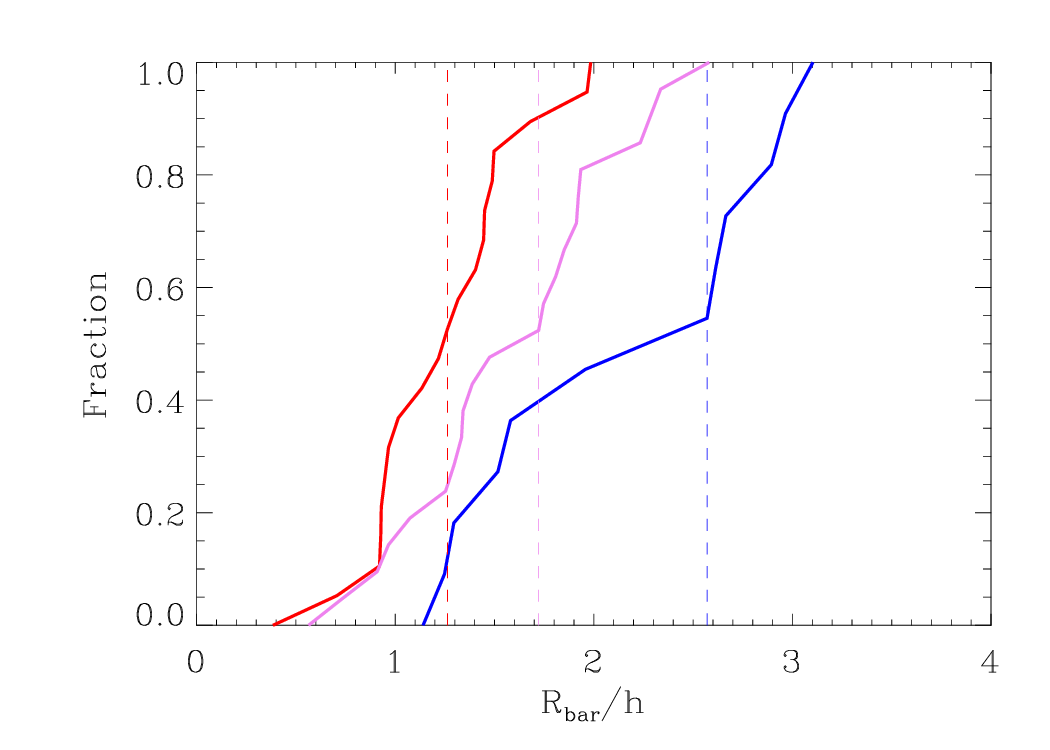}
    \caption{Bar radii scaled by the disk scale lengths for galaxies located in the cluster (red line), filament (violet line), and field (blue line) environments. Vertical dashed lines shows the median values for galaxies in the cluster (red), filament (violet) and field (blue). The vertical lines mark the median values of the scale bar radius per each galaxy sample.}
    \label{fig:rbar_h}
\end{figure}

Figure \ref{fig:rbar_h} shows the cumulative distributions of $R_{\mathrm{bar}}/h$ for the cluster, filament, and field galaxies. The figure indicates that scaled bars in the cluster are less prominent than those in the field, with filament galaxies exhibiting intermediate properties. Specifically, the median $R_{\mathrm{bar}}/h$ values for the cluster, filament, and field samples are $1.26 \pm 0.09$, $1.72 \pm 0.11$, and $2.57 \pm 0.21$, respectively (see also Table~\ref{tab:mean_value}). The KS and AD tests indicate that the three $R_{\mathrm{bar}}/h$ distributions are statistically different in shape (the AD test shows marginally similar distributions of scaled bar length when comparing bars from Virgo and filaments). In addition, the Welch t-test reveals that their median values also differ significantly. We therefore conclude that the $R_{\mathrm{bar}}/h$ distributions differ both in shape and in median across the three environments, with bars in the Virgo cluster being less prominent than those in the field (see Tab. \ref{tab:test_stat}).

\begin{table}[]
    \centering
    \caption{Median value and corresponding standard deviation of the bar and disk parameters for the galaxy samples from the different galaxy environments.}
    \begin{tabular}{lccc}
    \hline \hline
    Parameter & Virgo cluster & filaments & field \\
    \hline
    $\langle R_{\rm bar}\rangle$ [kpc] & $2.54\pm0.34$ & $3.29\pm0.38$ & $4.44\pm0.81$ \\
    %$\langle S_{\rm b}\rangle$ & $0.4\pm0.2$ & $0.5\pm0.2$ \\
    $\langle h \rangle$ [kpc] & $2.10\pm0.27$ & $2.17\pm0.35$ & $2.34\pm0.33$ \\
    %$\langle R_{\rm 90}\rangle$ [kpc] & $3.4\pm1.0$ & $9.3\pm4.0$ \\
    $\langle R_{\rm bar}/h \rangle$ & $1.26\pm0.09$ & $1.72\pm0.11$ & $2.57\pm0.21$ \\
    %$\langle R_{\rm b}/R_{\rm 90}\rangle$ & $0.6\pm0.3$ & $0.7\pm0.3$\\
    \hline
    \end{tabular}
    \tablefoot{Uncertainties of the median values correspond to $\sigma/\sqrt{N}$, where $\sigma$ and $N$ are the standard deviation and the number of galaxies.}
    \label{tab:mean_value}
\end{table}

\begin{table*}[]
    \centering
    \caption{Probability values for the statistical tests.}
    \begin{tabular}{lcccc}
    \hline
    \multicolumn{5}{c}{Disk scale length} \\
    \hline \hline
    Test & Virgo - filaments & Virgo - field & filaments - field & Virgo - filaments - field\\
    \hline
    $p_{KS}$  & 0.99  & 0.99  & 0.98  \\
    $p_{AD}$   & 0.25 & 0.25 & 0.25 & 0.25\\
    $p_{t}$ & 0.70  &  0.99 & 0.74 \\
    %$\langle R_{\rm b}/R_{\rm 90}\rangle$ & $0.6\pm0.3$ & $0.7\pm0.3$\\
    \hline \hline
    \multicolumn{5}{c}{Bar radius} \\
    \hline \hline
    Test & Virgo - filaments & Virgo - field & filaments - field & Virgo - filaments - field\\
    \hline
    $p_{KS}$ & 0.38  & 0.13  & 0.31  \\
    $p_{AD}$  & 0.24 & 0.02 & 0.17 & 0.04\\
    $p_{t}$ & 0.15  & 0.01 & 0.10 \\
    \hline \hline
    \multicolumn{5}{c}{Scaled bar radius} \\
    \hline \hline
    Test & Virgo - filaments & Virgo - field & filaments - field & Virgo - filaments - field\\
    \hline
    $p_{KS}$  & 0.05  & 0.005  & 0.05  \\
    $p_{AD}$  & 0.06 & 0.001 & 0.02 & 0.001\\
    $p_{t}$ & 0.03 & 0.0001& 0.02 \\
    \hline
    \end{tabular}
    \label{tab:test_stat}
\end{table*}

%\subsection{Bar properties in different environments}

%\subsubsection{Bar radius as a function of the environment}

%\subsubsection{Bar strength as a function of the environment}

%\subsubsection{Correlations among bar properties in filaments}

%\subsection{Bar properties in groups from different environments}

%Groups seem not causing preprocessing: bars in both cluster and filaments but in no groups host shorter bars. Bars have shortered becasue of the cluster-filaments environment. Bars from filaments but no group are more similar to those from the cluster (which are not in groups. I think the result is in agreement with the idea that the main driver of bar transformation is the gas stripping and/or tidal effects when entering to a dense environment (both cluster and filaments) while galaxies in group are able to maintain their gas content. 

%\subsection{Galaxy light concentration and bar parameters}

\section{Discussion}
\label{sec:discussion}

The formation of bars in disk galaxies can result from internal dynamical instabilities within the disk itself and/or external perturbations, such as gravitational interactions or minor mergers, which can significantly alter the galaxy’s potential \citep{Pettitt2018,Cavanagh2020}. {A barred galaxy observed at the present epoch has likely experienced a series of gravitational encounters, whose frequency and intensity are modulated by the density of its surrounding environment (e.g., being higher in clusters than in the field).} In addition to gravitational interactions, galaxies in high-density environments—or during their infall into a cluster—can be affected by various physical processes that impact their gas and stellar content, potentially altering their internal structure. This complex assembly history makes it difficult to disentangle whether the observed bar originated predominantly through secular evolution driven by internal processes or/and was triggered by external interactions.

\subsection{Influence of the environment on bar formation}

Several observational studies provide evidence that bar formation may be influenced by galaxy interactions, as reflected in the dependence of the bar fraction on the local environmental density. In particular, these works show that the fraction of barred galaxies increases in higher-density environments \citep[e.g.,][]{Thompson1981, MendezAbreu2012, Lin2014,Yoon2019}, suggesting that interactions or dynamical processes more frequent in dense regions, such as tidal perturbations or minor mergers, could play a significant role in enhancing bar formation. This trend supports the idea that the external environment can modulate the secular evolution of disks through gravitational effects that favor the development of bar structures. Although the analysis of the bar fraction provides valuable insights into the potential influence of the environment on bar formation, this quantity alone does not reveal the specific mechanisms that triggered the bar.  Disentangling this  requires complementary analyzes, such as studying the structural properties of bars, the dynamical state of disks, or the interaction history of galaxies.

In this work, we have analyzed the bar lengths of three samples of galaxies with similar luminosities and stellar colors but spanning different environments: the Virgo cluster, filaments building the Cosmic Web around Virgo, and the field. The main result is the detection of significant differences in both the physical bar radii and the prominence of bars across environments, from the highest density region (the Virgo cluster) to the lowest (the field). Our first observational result is that barred galaxies show different color distributions depending on the galaxy environment: they concentrate along the red sequence in clusters, while in filaments and the field they exhibit a bimodal distribution between the red sequence and the blue cloud (see Fig.~\ref{fig:color_mag}). {This simply reflects the well-known trend between color of the galaxies and environment \citep[see e.g., ][]{Hogg2004}.} After homogeneously selecting massive barred galaxies from the red sequence, we observe that bars in cluster galaxies are shorter and less prominent than those in field galaxies, with filament galaxies exhibiting intermediate properties. These findings, which should be confirmed with larger samples and extended to other cluster environments, {indicate that the properties of bars in massive disk galaxies—specifically their length and prominence—vary systematically with environment, suggesting that interactions and local density play a role in the evolution of bars in present-day disk galaxies.}

The results obtained in the present work partially agree with those reported in Paper I. In both studies, it was found that the physical sizes of bars in Virgo cluster galaxies are smaller than those in the selected field environments. However, while Paper I concluded that bar prominence does not depend on the environment, the present analysis shows that bars in cluster galaxies are significantly less prominent than those in field or filament galaxies. This discrepancy likely arises from differences in the field galaxy samples used in the two studies. In the present work, we have constructed a more homogeneous and carefully selected sample of field galaxies, minimizing observational biases in the structural properties of the galaxies. In contrast, the field in Paper I was assembled from a set of galaxies from the literature combined with barred galaxies from the CALIFA survey. This sample is far less homogeneous than our current field sample. Moreover, the CALIFA galaxies themselves are not an ideal field comparison sample, as their selection was subject to biases, especially a size requirement imposed during sample construction \citep[see][]{Walcher2014}. This size selection could explain the significant difference in galaxy sizes between the Virgo and field galaxies reported in Paper I. The resulting mismatch in galaxy sizes might have prevented us from detecting any trend on the scaled bar radius with environment in that study, contrary to the clear trend we observe in the present work.

\subsection{Influence of environment on the secular evolution of bars}

{Idealized numerical simulations of barred galaxies for disks in isolation show that} the bar formation process can be broadly divided into three main stages, each characterized by different timescales. The first two stages—the initial bar formation and the subsequent buckling phase—span approximately 1 to 1.5 Gyr each.  The influence of the environment on these initial formation phases of the bars have been analyzed in several numerical simulations. Thus, \citet{Zheng2025} explored the growth timescales of bars formed either through internal instabilities of their host galaxies or triggered externally by tidal perturbations, using a series of controlled N-body simulations. The authors showed that tidal interactions can both promote or delay the onset of bar formation by advancing or postponing its emergence. However, once the bar begins to grow, its growth rate remains largely unchanged and comparable to that of bars formed through purely internal instabilities. Indeed, even in interaction scenarios, the subsequent evolution of the bar is primarily driven by the internal dynamical properties of the disk, which are largely unaffected by the external perturbation. Specifically, if the disk is prone to internal instabilities, these dominate the bar’s formation and evolution regardless of the tidal interaction. In contrast, in dynamically hot disks, tidal perturbations can trigger bar formation with a slower, linear growth of the structure. Similar results have been reported in other simulations, such as those by \citet{MartinezValpuesta2017} and \citet{Moetazedian2017}.

The third phase in the bar evolution is the so-called secular evolution, which takes several Gyrs of the lifetime of the galaxies. Therefore, the bars we observe today are most likely in their secular evolution stage. During this period, the bar exchanges angular momentum with other galactic components, such as the stellar disk and the dark matter halo. This exchange of angular momentum leads the bar to evolve, typically becoming longer in radius and rotating slower in its pattern speed \citep[see][]{Debattista2000, Athanassoula2002, Athanassoula2013}. This means that bars become increasingly prominent within the disk as they evolve during the secular evolution phase. The fact that we observe less prominent and shorter bars in the cluster environment compared to those in the field could indicate that the physical processes more frequent in high-density environments either slow down the secular evolution of the bar or lead to the formation of bars with intrinsically different properties. Can we identify which of the processes occurring in high-density environments is primarily responsible for slowing down the secular evolution of bars? Is this process driven solely by gravitational interactions, or is it the result of a combination of different physical mechanisms?

%Galaxies located in high-density environments experience different physical processes compared to those in the field. Among the mechanisms that could explain the slowdown of bar secular evolution are gas stripping, strangulation, and galaxy harassment. Gas stripping \citep{Gun1972, Quilis2000} and strangulation \citep{Larson1980} both lead to the depletion of a galaxy’s cold or warm gas, effectively quenching its star formation. This lack of gas prevents the formation of new stars in the bar region, thereby slowing its growth. Additionally, galaxies in dense environments can undergo repeated high-speed encounters with other galaxies—a process known as galaxy harassment \citep[][]{Moore1999}. These encounters can heat the galactic disk and increase its vertical instability, reducing the efficiency of the bar in redistributing angular momentum. As a consequence, the growth of the bar can be significantly slowed \citep[][]{Athanassoula2003, Athanassoula2013}.

Galaxies located in clusters can experience strong gravitatinonal interactions with the cluster potential \citep{Cuomo2022,smith2022}. N-body models have been used to simulate the strong interactions experienced by galaxies during their infall into clusters. The properties of the bars formed in these simulations arise solely from gravitational encounters between galaxies and/or the cluster potential, as gas physics is not included in such models. In particular, \citet{Aguerri2009} conducted a series of N-body simulations of massive disc galaxies to study the remnants produced by strong gravitational encounters, such as those occurring in cluster environments. In all cases, galaxies experienced significant stellar and total mass loss. Systems with small initial bulges and prograde orbits developed bar-like structures in their discs. The authors measured the ratio $R_{\mathrm{bar}}/h$ for these bars and found that it varied with the number of interactions each galaxy underwent. Specifically, $R_{\mathrm{bar}}/h$ values ranged from 1 to 3.5, with most cases showing $R_{\mathrm{bar}}/h > 1.5$ (see their Fig.~14), and only a few falling between 1 and 1.5. We can compare these simulated values with those measured in our sample of Virgo cluster galaxies. In our case, only a small fraction (about 15\%; see Fig.~\ref{fig:rbar_h}) of Virgo galaxies exhibit $R_{\mathrm{bar}}/h > 1.5$. This suggests that the bars observed in Virgo are unlikely to be the result of strong gravitational interactions alone. Similarly, \citet{Lokas2016} explored the formation of bars in Milky Way-like galaxies falling into a Virgo-like cluster using N-body simulations. The author found that galaxies on tighter orbits develop longer and stronger bars at the end of the simulation, concluding that tidal interactions between the stellar disc and the cluster potential enhance bar formation.

Taken together, both sets of N-body simulations predict that bars induced purely by gravitational interactions should be relatively large and prominent in the remnant disks. However, the observational results presented here reveal the opposite trend: bars in Virgo cluster galaxies are, on average, shorter and less prominent than those in field galaxies. This discrepancy suggests that, although tidal interactions can indeed trigger or strengthen bars, real galaxies falling into clusters like Virgo are subject to additional environmental mechanisms. Processes such as harassment \citep[][]{Moore1996, moore1999}, ram-pressure stripping \citep{Gun1972, Quilis2000}, and starvation \citep{Larson1980} can suppress gas inflow, heat the stellar disk, and alter the secular evolution of bars---ultimately resulting in shorter and weaker bars than predicted by simulations that consider only gravitational effects.

%\textcolor{red}{A larger fraction of filament galaxies ($\sim40\%$) presents a $R_{\mathrm{bar}}/h$ larger than 1.5, showing that the influence of interactions can be different in this less dense environment.} Moreover, it is important to note that the remnants in the \citet{Aguerri2009} simulations evolve into dwarf systems as a consequence of the severe mass loss induced by the interactions. These bar-bearing dwarf galaxies are fundamentally different from our sample of massive galaxies, further supporting the idea that strong interactions are not the primary drivers of bar formation in our Virgo cluster sample of massive galaxies. Although it is beyond the scope of this work, it would be interesting in future studies to measure the $R_{\mathrm{bar}}/h$ ratio for dwarf galaxies in the Virgo cluster and compare these values with those predicted by the simulations. \textcolor{red}{Can we discuss this paragraph? I think it is interesting but we cannot do a fair comparison since in 2009 you analysed dwarf galaxies, therefore, I am worried that a comparison can be misleading and actually suggest oppostie conclusions wrt what we want to say.}

\citet{Zana2018} analyzed the formation of bar structures in a series of cosmological simulations of Milky Way-like galaxies. This type of simulation includes additional physics, modeling the evolution of galaxies in a more realistic way than pure N-body simulations.  In particular, they studied the bar properties in a set of simulations in which they altered the history of galaxy-satellite interactions at a time when the main galaxy had not yet developed any non-axisymmetric bar-like structure. They found that the main effect of the environment is to delay the time of bar formation, but these interactions do not significantly alter the global properties of the bar, such as its radius. They measured the ratio $R_{\mathrm{bar}}/h$ in their simulations, obtaining values between 1.2 and 2.0 in the different evolutionary scenarios. In this context, larger values of $R_{\mathrm{bar}}/h$ were found in simulations with fewer satellite mergers and fly-bys, while the galaxy with the original evolutionary history—including multiple late mergers and fly-by interactions—showed the smallest $R_{\mathrm{bar}}/h$ ratio.

These results indicate that $R_{\mathrm{bar}}/h$ depends on the environmental interaction history of the galaxy: galaxies experiencing frequent interactions develop less prominent bars. This trend is in agreement with our observational findings, where galaxies in the field exhibit larger $R_{\mathrm{bar}}/h$ ratios than those in the cluster environment. As suggested by the simulations, this supports the idea that high-density environments, where interactions are more frequent, can delay or slow down the secular evolution of bars. {However, the differences observed in bar properties between pure N-body and cosmological simulations are consistent with the possibility that bar evolution is influenced by processes beyond gravitational interactions, or that cosmological simulations offer a more complete representation of a galaxy's lifetime evolution than idealized experiments with only a few orbital passages.}

\subsection{Pre-processing of bars in filaments}

It is worth noting that the residence time of galaxies in clusters can span several Gyr \citep{Haines2015, Jaffe2016}, which raises the question of whether the observed bars were transformed before or after the galaxy entered the cluster. This temporal dimension is crucial, as it could imply that the current bar properties reflect either a legacy of evolution in the field or the prolonged influence of the dense environment.

Galaxies currently located in clusters may have evolved through different environments over their lifetimes, not necessarily matching their present-day surroundings. These varying environmental conditions experienced throughout a galaxy’s history can significantly influence its properties. Using high-resolution hydrodynamic zoom-in simulations, \citet{Han2018} explored the impact of environments encountered prior to cluster infall, a process known as pre-processing. They found that nearly 50$\%$ of present-day cluster members were satellites of other host halos before joining the cluster. Similarly, \citet{Joshi2017} investigated the tidal stripping of cluster galaxies and found that, on average, galaxies that were previously group members lost 35–45$\%$ of their peak mass before falling into the cluster, indicating a strong influence of pre-processing on these systems. Moreover, they showed that galaxies falling into the cluster as isolated halos quickly “catch up” in terms of mass loss, reaching levels comparable to those that experienced pre-processing in groups.

Our sample includes barred galaxies located in the filamentary infall regions of the Virgo cluster. The physical radii of bars in the filament environment are larger than those in the cluster ($3.29 \pm 0.38$ kpc vs. $2.54 \pm 0.34$ kpc), although different statistical tests do not indicate that this difference is statistically significant. In contrast, the tests do reveal a statistically significant difference between cluster and filament galaxies when bar sizes are normalized by the disk scale length, indicating that bars in filament galaxies are more prominent than those in cluster galaxies. A similar result is found when comparing both the physical and scaled bar lengths of galaxies in the field and in filaments. This suggests that environmental pre-processing in filaments---occurring before galaxies enter the cluster---may already start to influence bar growth and prominence, highlighting the importance of filamentary structures in the evolutionary path of disk galaxies.

\subsection{Alternative explanations: internal drivers of bar evolution}

As shown in the previous subsections, our observational results can be explained by the various processes that galaxies undergo during their lifetime across different environments. However, numerical simulations have demonstrated that bar formation and evolution is a complex process that depends on several internal galaxy parameters, which influence the evolution of galactic disks \citep[see, e.g.,][]{Athanassoula2013}. Therefore, it is worth considering whether internal properties related to bar formation and evolution---such as central mass concentration, dynamical friction, or initial gas fraction---could also play a role in explaining the results presented in this work.

%Numerical simulations have shown that bar formation is a complex process depending on many different galaxy parameters that play a role in the evolution of galaxy disks \citep[see e.g][]{Athanassoula2013}.

The presence of a large concentration of central mass in galaxies (e.g., massive bulges, pseudo-bulges or central black holes) can inhibit or limit bar growth by stabilizing the inner disk \citep{Norman1996, Shen2004, Athanassoula2005}. However, studies such as \citet{Gadotti2011} show that field galaxies with pseudo-bulges can still host bars as large as those in field galaxies without pseudo-bulges. This indicates that central mass concentration alone cannot account for the shorter bars observed in cluster environments \citep[see also][]{Erwin2005,Laurikainen2007}.

The dynamical friction exerted by the dark matter halo on the bar is a key mechanism that could contribute to slowing down the secular evolution of bars in galaxies residing in dense environments. According to \citet{Debattista2000} and \citet{Athanassoula2003}, the exchange of angular momentum between the stellar bar and the surrounding dark matter halo can lead to a significant braking of the bar’s pattern speed. In environments where galaxies host more massive or centrally concentrated dark matter halos, this friction becomes more efficient, absorbing angular momentum from the bar more rapidly and thus hindering its growth or even weakening it over time. Simulations predict that galaxies destined to reside in massive clusters collapse earlier and therefore develop more concentrated dark matter halos than their field counterparts at a given stellar mass \citep[][]{Bullock2001, Wechsler2002, Gao2005}. These denser halos in galaxies in clusters could partly explain why bars in high-density environments seem to evolve more slowly compared to those in the field.

Additionally, variations in the initial gas fraction of galaxies may play an important role in regulating bar evolution \citep[see e.g.][]{Athanassoula2013}. Galaxies formed or processed in high-density regions tend to exhibit lower cold gas reservoirs due to earlier star formation or environmental processes such as preprocessing in groups. This scarcity of gas limits the inflow of material along the bar, reducing the fuel available for new star formation and, consequently, slowing down the bar's secular growth. Furthermore, the chemical enrichment of galaxies in dense environments can proceed more rapidly, altering the gas cooling properties and further impacting the ability of the bar to evolve efficiently \citep{Cooper2008}.

We can conclude that either the large central density of dark matter halos or the low initial gas fraction of galaxies in clusters could partially explain the observations shown in the present work. However, this goes beyond the scope of this paper: we suggest these mechanisms to be explored in future observational or theoretical works.

\subsection{Sample limitations}

Our sample of galaxies presented in this work has some limitations that should be considered. It is important to note that our sample is limited to luminous galaxies with stellar masses above $10^{10} M_\odot$, which excludes potential populations of lower-mass galaxies that could show different trends in their bar properties. In addition, only galaxies within the red sequence has been considered. Blue (late-type) galaxies could show different properties on their bars that should also be explored in the different explored environments.

The number of galaxies in our sample is quite limited specially among  the galaxies in the field (only 12), due to the fact that most of barred galaxies from the field are significantly detached from the red sequence and generally present a intermediate to low masses (see Fig.~\ref{fig:color_mag}). Indeed, we attempted to expand the field galaxy sample by varying the local galaxy overdensity $\delta_g$ threshold, but had to find a compromise to avoid strong contamination between field and filament galaxies, which present overlapping cumulative $\delta_g$ distributions (see Fig.~\ref{fig:overdens_SDSS}). It would be needed to increase the number of galaxies specially in the field to confirm statistically some of the results as the dependence of the bar radius with the galaxy environment. 

%In our previous work \citep[][]{Aguerri2023}, we found that bars in Virgo cluster galaxies were significantly shorter than those in the corresponding field sample. Specifically, we measured a mean bar radius of $R_{\mathrm{bar}} = 2.6 \pm 1.5$ kpc for the Virgo galaxies, a value very similar to that obtained in the present study. However, the physical bar size in their field sample was $R_{\mathrm{bar}} = 6.1 \pm 3.0$ kpc, notably larger than the mean value measured for our current field galaxies (see Table~\ref{tab:mean_value}).

%These differences in the bar lengths of the field samples likely reflect variations in the selection criteria for field galaxies between \citet{Aguerri2023} and this study. In particular, in \citet{Aguerri2023}, the field sample was assembled from a heterogeneous set of galaxies drawn from the literature, combined with barred galaxies from the CALIFA survey. This sample is far less homogeneous than our current field sample. Moreover, the CALIFA galaxies themselves are not an ideal field comparison sample, as their selection was subject to biases---especially a size requirement imposed during sample construction \citep[see][]{Walcher2014}.

%This size selection could explain the significant difference in bar and galaxy sizes between Virgo and field galaxies reported in \citet{Aguerri2023}. The resulting mismatch in galaxy sizes might have prevented us from detecting any trend of $R_{\mathrm{bar}}/h$ with environment in that study, contrary to the clear trend we observe in the present work.

Another limitation of our sample is that we have only analyzed one cluster. A natural next step to expand this study would be to analyze  other local and intermediate-redshift clusters will allow us to determine whether the trends observed in Virgo are universal or specific to this cluster. Some results have been obtained in other environments. For example, \citet{Marinova2009} and \citet{Lansbury2014} found in Coma and A901/902 that the frequency and prominence of bars decrease in regions of higher local density, consistent with our findings in Virgo. Similarly, \citet{Barazza2009} analyzed clusters at $0.4 < z < 0.8$ and found a decline in the fraction of barred galaxies toward the cluster centers. These results suggest that the trend we observe in Virgo could be a common feature in both local and intermediate-redshift clusters. Similar considerations can be driven for the filament environment as well, as the considered filaments differ in properties such as extension, galaxy density, distance from the cluster, and so on \citep{Lee2021}, while our study is limited to those around the Virgo cluster. Therefore, a more exhaustive selection of filaments could be desirable for further studies. 

\section{Conclusions}
\label{sec:conclusions}

We have carefully selected and analyzed a sample of 54 barred, red, and luminous galaxies located in three different environments: the Virgo cluster (20), the filaments around the Virgo cluster (22), and the field (12). For these galaxies, we obtained two main structural parameters: their bar radii and disk scale lengths.

We have compared the cumulative distributions of these parameters in different environments and driven statistical tests to highlight similarities or differences. The results of the KS tests show that the disk scale lengths of galaxies in the three environments are statistically similar. This indicates that the galaxies have similar disks. Therefore, any differences in bar properties must arise from environmental effects. 

As a result, the median bar radii differ among the three environments: bars in cluster galaxies are shorter than those in filament or field galaxies ($2.54\pm0.34$ kpc for cluster, $3.29\pm0.38$ kpc for filament $4.44\pm0.81$ kpc for field environments). While the KS test does not reveal strong differences, the AD and Welch t-tests do reveal statistically significant differences between the Virgo and field bar samples. 

On the other hand, the prominence of bars relative to their disks does show clear environmental trends being the median values $1.26\pm0.09$, $1.72\pm0.11$, and $2.57\pm0.21$ for bars in cluster, filament, and field environments, respectively.  This indicates that bars in high-density environments are significantly less prominent than those in filaments and the field, and in this case the different statistical tests run confirm the statistical differences between the three distributions. 

The fact that bars in the highest density environments are both shorter and less prominent suggests that the secular evolution of bars is slowed down in galaxies residing in hostile environments. Interactions between galaxies, as well as with the cluster potential, can hinder the growth and evolution of bars in cluster galaxies. Furthermore, this slowing-down process may begin already in filaments, before galaxies fully enter the cluster region, since we find that bars in filament galaxies are significantly less prominent than those in the field.

These results highlight the crucial role of environment in regulating the secular evolution of bars in massive and red galaxies. {This suggests that bars may provide useful indications of past and ongoing environmental influences on galaxies.}  Future studies combining deeper observations and high-resolution cosmological simulations will be essential to fully unravel the interplay between bar evolution and environment across cosmic time and different environments.

\begin{acknowledgements} We thank the referee for their helpful comments, which have contributed to improving the quality of the paper. VC acknowledges the support provided by ANID through the FONDECYT grants no. 3220206 and 11250723. JALA acknowledges support from the Agencia Estatal de Investigación del Ministerio de Ciencia, Innovación y Universidades (MCIU/AEI) under grant “WEAVE: EXPLORING THE COSMIC ORIGINAL SYMPHONY, FROM STARS TO GALAXY CLUSTERS” and the European Regional Development Fund (ERDF) with reference PID2023-153342NB-I00 / 10.13039/501100011033. VC thank Universidad de Atacama for the hospitality during the preparation of this paper. LM and JALA acknowledges a support grant from the Joint Committee ESO-Government of Chile (ORP 058/2023). NCC acknowledges a support grant from the Joint Committee ESO-Government of Chile (ORP 028/2020) and the support from FONDECYT Postdoctorado 2024, project number 3240528. SZ acknowledges the financial support provided by the Governments of Spain and Arag\'on through their general budgets and the Fondo de Inversiones de Teruel, the Aragonese Government through the Research Group E16-23R, and the Spanish Ministry of Science and Innovation and the European Union - NextGenerationEU through the Recovery and Resilience Facility project ICTS-MRR-2021-03-CEFCA. We thank R. Smith, Y. Rosas-Guevara, B. Cervantes Sodi and K. Chim for fruitful discussions. We acknowledge that the location where this work took place and where the data were collected lies on Indigenous lands and skies. We recognize and respect the enduring relationship that Indigenous people have with these lands and skies.

\end{acknowledgements}


\bibliographystyle{aa}
\bibliography{biblio} 

\end{document}